\documentclass[twocolumn,superscriptaddress,floatfix,prb]{revtex4-1}
\usepackage{graphicx,amsfonts,amssymb,amsmath,hyperref,hypcap,enumerate}

\newcommand{\ba}{\boldsymbol{a}}
\newcommand{\ee}{\boldsymbol{e}}
\newcommand{\pp}{\boldsymbol{p}}
\newcommand{\qq}{\boldsymbol{q}}
\newcommand{\QQ}{\boldsymbol{Q}}
\newcommand{\RR}{\boldsymbol{R}}
\newcommand{\rr}{\boldsymbol{r}}
\newcommand{\KK}{\boldsymbol{K}}
\newcommand{\kk}{\boldsymbol{k}}
\newcommand{\dd}{\boldsymbol{d}}
\newcommand{\DD}{\boldsymbol{D}}
\newcommand{\bb}{\boldsymbol{b}}
\newcommand{\jj}{\boldsymbol{j}}
\newcommand{\JJ}{\boldsymbol{J}}
\newcommand{\GG}{\boldsymbol{G}}

\begin{document}
\title{Theory of optical absorption by interlayer excitons \\ in transition metal dichalcogenide
heterobilayers}

\author{Fengcheng Wu}
\affiliation{Materials Science Division, Argonne National Laboratory, Argonne, Illinois 60439, USA}

\author{Timothy Lovorn}
\affiliation{Department of Physics, University of Texas at Austin, Austin, Texas 78712, USA}

\author{A. H. MacDonald}
\affiliation{Department of Physics, University of Texas at Austin, Austin, Texas 78712, USA}

\date{\today}

\begin{abstract}

We present a theory of optical absorption by interlayer excitons in a heterobilayer
formed from transition metal dichalcogenides.  The theory accounts for the presence of small relative 
rotations 
that produce a momentum shift between 
electron and hole bands located in different layers, and a moir\'e pattern in real space.
Because of the momentum shift, the optically active interlayer excitons are located at the 
moir\'e Brillouin zone's corners, instead of at its center, and would have 
elliptical optical selection rules if the individual layers were translationally invariant.
We show that the exciton moir\'e potential energy  restores circular optical selection rules
by coupling excitons with different center of mass momenta.  
A variety of interlayer excitons with both senses of circular optical activity, 
and energies that are tunable by twist angle, are present at each valley.  
The lowest energy exciton states are generally localized near the exciton potential energy 
minima.  We discuss the possibility of using the moir\'e pattern to achieve scalable
two-dimensional arrays of nearly identical quantum dots.

\end{abstract}

\maketitle

\section{Introduction} 
Motivated by the discovery of unprecedented optical \cite{Xu_Review} and electronic \cite{Mak_NbSe2, Cobden_WTe2} properties, the monolayer transition metal dichalcogenides (TMD) class of 
two-dimensional materials  is currently under active study.  
Because of strong Coulomb interactions between electrons and holes, the optical absorption of monolayer TMD semiconductors like MoS$_2$ is dominated by excitonic features.\cite{ye2014, He2014, chernikov2014, Qiu2013}
In TMDs excitons have two-fold valley degeneracy because the material's 
band extrema are located at inequivalent momenta at the triangular lattice
Brillouin zone corners that are related by time-reversal symmetry.  
Valley dependent optical selection rules enable optical 
manipulation of the valley degree of freedom, \cite{cao2012valley, zeng2012valley, mak2012control, Di2012}
providing access to many interesting phenomena, including light-induced Hall effects, \cite{mak2014valley}
a valley-selective optical Stark effect,\cite{kim2014, sie2015} and a large valley-exclusive Bloch-Siegert shift\cite{Sie1066}.

Van der Waals heterobilayers composed of different monolayer TMDs provide an 
even richer platform in which new properties can be engineered.
WX$_2$/MoX$_2$ (X= S, Se) heterobilayers have been experimentally 
realized \cite{Hong2014, gong2014, gong2015, liu2014evolution, rivera2016valley}, and
have a type-II band alignment in which the conduction and valence band edges are associated with MoX$_2$ and WX$_2$ respectively. The lowest-energy excitons are therefore spatially indirect since the 
constituent electrons and holes are located primarily in different layers.
The interlayer excitons have a long lifetime and  an electrically tunable energy because of their spatially indirect nature\cite{Rivera2015}, but still possess a sizable electron-hole binding energy. 
These features make TMD bilayers as a strong candidate system to realize spatially indirect exciton 
condensation.\cite{fogler2014high, Wu2015}

Heterobilayers of two-dimensional materials
exhibit long period moir\'e patterns when they have a small lattice constant mismatch and/or 
relative orientation angle.  
Moir\'e patterns can induce dramatic changes in material properties that can be 
altered by controlling twist 
angle. In graphene-based heterostructures, the moir\'e pattern strongly modifies the 
Dirac spectrum,\cite{hunt2013, Wang2016, yankowitz2012, ponomarenko2013} leads to 
Hofstadter-butterfly spectra in a strong magnetic field,\cite{hunt2013, dean2013, Kim_bilayer} and can 
assist optical spectroscopy in the fractional quantum Hall regime.\cite{Wu_MacDonald} 
In TMD heterobilayers, the possibilities are richer because of the wider 
variety of materials, and they are still relatively unexplored.
In a previous paper\cite{Wu2017} we have shown that a
moir\'e pattern profoundly alters the spectrum of optical absorption due to 
{\it intralayer} excitons by producing satellite excitonic peaks,
and enables the design of topological bands of {\it intralayer} excitons.
Moir\'e pattern in TMDs can also be used to engineer electronic topological insulator nano-dots and nano-stripes\cite{Tong2017}.

A comprehensive theory of interlayer excitons in moir\'e pattern is desirable for interpreting ongoing and future experiments.
In a pioneering theoretical paper \cite{Yu2015} by Yu {\it et al}., it was shown that optically 
active interlayer excitons are located at the corners of the moir\'e Brillouin zone (MBZ) 
instead of its center, and that they have elliptical optical selection rules instead of circular.
These predictions are however based on a theory that omits 
one important ingredient, namely the spatial modulation of exciton energy produced
by the moir\'e pattern. In this paper we present a theory of interlayer excitons in a moir\'e pattern
that does take account of the exciton potential energy, which has a sizable magnitude according to 
our {\it ab initio} calculations.  We explain why this potential energy plays an essential role in determining 
exciton properties.

Our theory predicts that the exciton potential energy has two major effects on interlayer excitons.
First, the periodic exciton potential inevitably mixes and splits the three independent exciton 
center-of-mass eigenstates that are otherwise degenerate at the 
MBZ corners.  The new exciton eigenstates respect the 
rotational symmetries of the moir{\' e} pattern and for this reason circular optical selection rules
must be restored.  Mixing of center of mass eigenstates by the exciton potential 
energy splits the spectral weight associated with interlayer excitons across 
a variety of distinct states whose energy splittings can be tuned by adjusting twist angles. 
Interlayer excitons also have a valley degree of freedom, but the optical 
selection rules are not locked to valley.  In particular, different interlayer 
excitons at a given valley can absorb light with different circular polarization.  
This can have important experimental consequences, as we will explain.
Second, the moir\'e potential generally results in localized exciton states that are confined near 
the potential minima. This suggests a Van der Waals heterostructure
moir\'e based strategy for realizing two-dimensional arrays of nearly identical quantum dots with nanoscale.
When the moir\'e potential is fine tuned such that it has an enhanced six-fold rotational symmetry, 
the exciton moir\'e band structure has Dirac dispersion within the light cone.  

Our paper is organized as follows. In Secs.~\ref{Sec:Theory} and \ref{Sec:OME}, we formulate the optical matrix element for interlayer excitons. The derivation closely follows that in Ref.~\onlinecite{Yu2015}. In Sec.~\ref{Sec:Potential}, the exciton moir\'e potential energy is constructed. In Sec.~\ref{Sec:Oabsorb}, we present our main results on the optical absorption spectrum for interlayer excitons in moir\'e pattern. In Sec.~\ref{Sec:Dirac}, we show the presence of Dirac cones in the exciton band structure when the moir\'e potential is fine tuned. Finally in Sec.~\ref{Sec:Disc}, experimental implications of our work are discussed.

\section{Theory of Moir\'e Heterobilayer Interlayer Excitons} 
\label{Sec:Theory}
We study WX$_2$/MoX$_2$ (X= S, Se)  
TMD heterobilayers with a small relative twist angle $\theta$.
We choose heterobilayers with a common chalcogen (X) atom, 
because these have small lattice constant mismatches ( $\delta \sim 0.1$\%).
The moir\'e periodicity $a_M$ is given by:
\begin{equation}
a_M \approx a_0/\sqrt{\theta^2+\delta^2},
\end{equation}
where the lattice constant mismatch $\delta$ is defined as $|a_0-a_0'|/a_0$, and 
$a_0$ and $a_0'$ are the lattice constants of the two layers.
At twist angles larger than $\sim 0.1^{\circ}$, we can neglect $\delta$.
To simplify the discussion, we set $\delta$ to be zero below.
The low-energy conduction band states of WX$_2$/MoX$_2$ bilayers 
are localized in MoX$_2$,
and the low-energy valence band holes are localized in the WX$_2$ layers.
Mixing between layers is weak because of the Van der Waals character of the 
heterobilayer, and because of the heterobilayer band alignment.\cite{CZhang2017}

\begin{figure}[t!]
	\includegraphics[width=1\columnwidth]{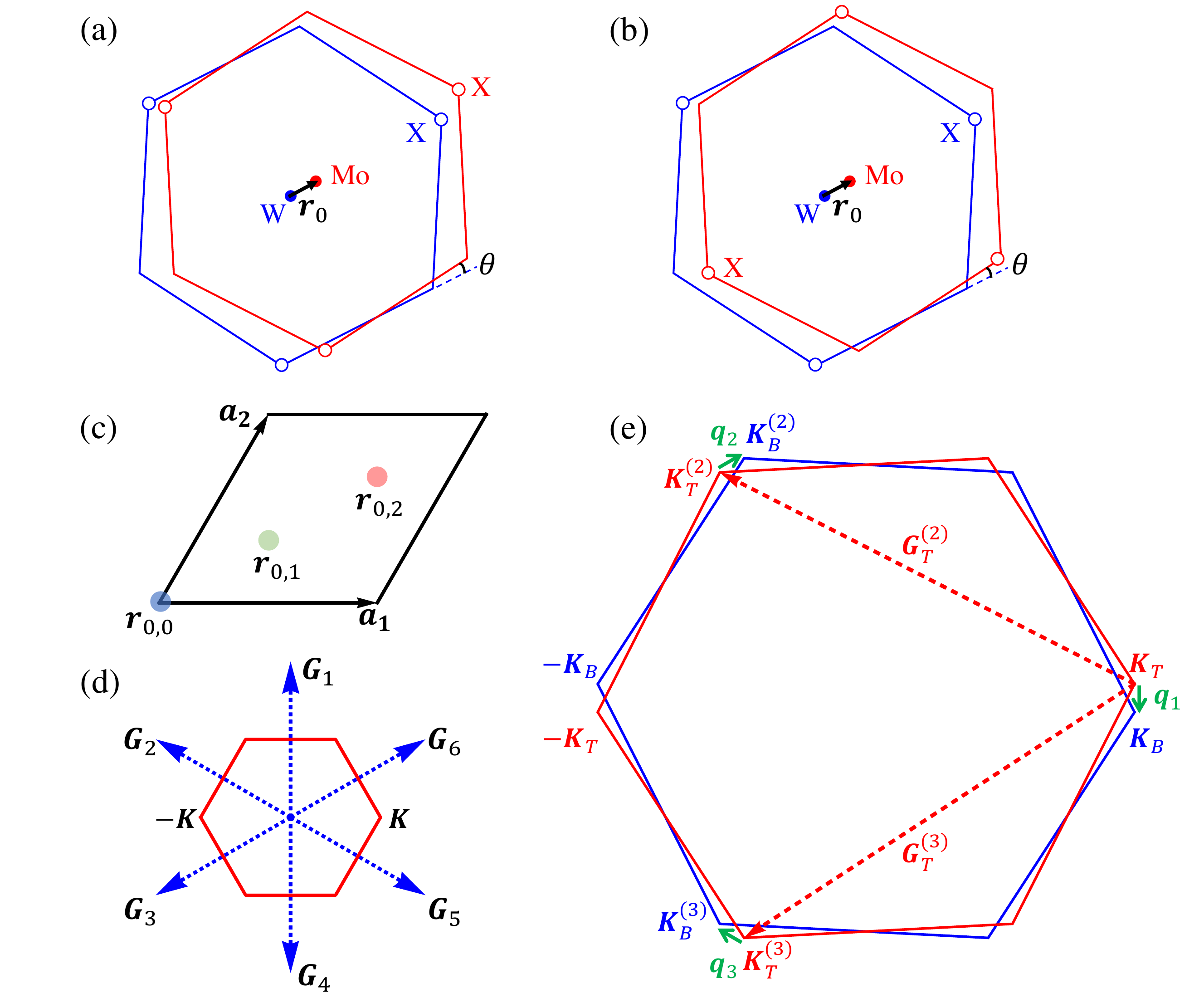}
	\caption{ (a) 
	Illustration of AA stacking with a small twist angle $\theta$ and an in-plane displacement $\rr_0$.
	The blue and red hexagons mark the Wigner-Seitz cells of the two layers.
	(b) The corresponding illustration of AB stacking. AA and AB stacking are distinguished by a rotation of the top layer (MoX$_2$) by $\pi$ around the metal (Mo) axis.   (c) Unit cell of the bilayer when $\theta=0$.  The vectors $\rr_{0,n}$ in this figure are high symmetry points as discussed in the main text.    (d) Brillouin zone and the first shell reciprocal lattice vectors for the  bilayer with $\theta=0$. (e) Brillouin zones associated with the top and bottom layers in a twisted bilayer. }
	\label{Fig:Pattern}
\end{figure}

The bilayer has two distinct stacking orders AA and AB, which are
illustrated in Figs.~\ref{Fig:Pattern}(a) and \ref{Fig:Pattern}(b).
Both configurations are experimentally relevant. \cite{rivera2016valley, wilson2016band}
Our choice of stacking convention allows us to study interlayer excitons in 
AA and AB stacking systems in a unified manner.

We consider interlayer excitons composed of electrons (holes) 
residing in the $\pm \KK_T$ ($\pm \KK_B$) valleys of a top (T) MoX$_2$ [bottom (B) WX$_2$] layer. 
The electron-hole exchange interaction, which couples intralayer   
$+ \KK$ and $- \KK$ valley excitons\cite{yu2014dirac, Glazov2014, Yu2014, wu2015Exciton}, is extremely weak
for interlayer excitons because the electrons and holes are spatially separated. 
We therefore neglect the exchange interaction
and study $\pm \KK$ valley excitons separately.  
Since excitons in opposite valleys are time-reversal partners, 
we focus on the $+\KK$ valley excitons unless otherwise stated. 
When the weak hybridization 
between layers is neglected, exciton states can 
be labeled by center-of-mass momentum $\QQ$. 
The corresponding exciton wave function is:
\begin{equation}
|\QQ\rangle=\frac{1}{\sqrt{\mathcal{A}}}\sum_{\kk}f_{\kk} a_{c (\KK_T+\kk+\frac{m_e}{M} \QQ)}^\dagger a_{v(\KK_B+\kk-\frac{m_h}{M} \QQ)} |G\rangle,
\label{ExWF}
\end{equation}
where $|G\rangle$ is the ground state of the heterobilayer,
and $a_{c (\KK_T+\qq)}^\dagger$ [$a_{v (\KK_B+\qq')}$] creates an electron (a hole) in the  $\KK$-valley conduction (valence) band of the MoX$_2$ (WX$_2$) layer.
The meaning of various quantities that enter (\ref{ExWF}) is as follows: 
$\mathcal{A}$ is the system area, 
 $\kk$ is the electron-hole relative momentum,
and $\QQ$ is the electron-hole center-of-mass momentum.
In making these definitions we have adopted the usual convention of 
measuring momentum from the band extrema points in each layer.   
In Eq.~(\ref{ExWF}) $m_e$ and $m_h$ are respectively the 
electron and hole effective masses, $M=m_e+m_h$ is the exciton total mass, and 
$f_{\kk}$ is the electron-hole relative-motion wave function in momentum space.  
We only study the lowest-energy interlayer excitons, for which $f_{\kk}$ has 
$s$-wave symmetry, {\it i.e.} it depends only on the magnitude of $\kk$.
$f_{\kk}$ is normalized such that $\langle \QQ |\QQ\rangle=1$.

When the bilayer has zero twist angle
the optically active excitons are located at $\QQ=0$ because light carries a vanishingly small momentum. 
At a finite twist angle, the Brillouin zones associated with the top and bottom 
layers have a relative rotation, leading to a finite momentum displacement between electron and hole 
bands. [See Fig.~\ref{Fig:Pattern}(e)]   When we adopt the standard and convenient convention of 
measuring momentum in each layer from its Brillouin-zone corner band extremum,
the optically active excitons in the
$\KK$ valley are shifted to momenta 
\footnote{If we choose $\QQ+\KK_T-\KK_B$ as the center-of-mass momentum, then optically active excitons are located at zero momentum (up to moir\'e reciprocal lattice vectors).} 
\begin{equation}
\QQ=(\KK_B+\GG_B)-(\KK_T+\GG_T),
\label{conservation_law}
\end{equation}
where $\GG_B$ and $\GG_T$ are the reciprocal lattice vectors of the bottom and top layers respectively.
By time-reversal symmetry, bright excitons at the $-\KK$ valley are located at 
momenta opposite to those in (\ref{conservation_law}). We
prove Eq. (\ref{conservation_law}) below.  

The interlayer optical matrix element between the ground state and the exciton state  is:
\begin{equation}
\JJ(\theta,\rr_0,\QQ)=\frac{1}{\sqrt{\mathcal{A}}} \langle G | \hat{\jj} |\QQ\rangle,
\end{equation}
where $\hat{\jj}$ is the current operator.
We evaluate $\JJ$
using a local approximation valid for long moir\'e periods.  
Because the two layers are weakly coupled and have separate lattice translational symmetry, 
the conduction band electron and valence band
hole creation operators at a given momentum can be expanded in terms 
of their Wannier function counterparts: 
\begin{equation}
\begin{aligned}
a_{v (\KK_B+\qq)}&=\frac{1}{\sqrt{N}}\sum_{\RR} e^{-i (\KK_B+\qq)\cdot \RR} a_{v \RR},\\
a_{c (\KK_T+\qq')}^\dagger&=\frac{1}{\sqrt{N}}\sum_{\RR'} e^{i(\KK_T+\qq')\cdot (\RR'+\rr_0)} a_{c (\RR'+\rr_0)}^\dagger,
\end{aligned}
\label{Bloch}
\end{equation}
where $N$ is the number of unit cells in each layer, $c$ and $v$ are the Wannier labels 
identified with the conduction and valence bands, and $\RR$ is a triangular lattice Bravais vector.
We assume 
a two-center approximation\cite{Bistritzer2011,Yu2015} for the optical 
matrix element and for its dependence on a rigid displacement by $\rr_0$ of one layer with 
respect to the other.  That is to say we assume that 
\begin{equation}
\begin{aligned}
 &\langle G |  \hat{\jj} a_{c (\RR'+\rr_0)}^\dagger a_{v \RR}  |G \rangle\\
\approx &\jj(\RR'+\rr_0-\RR)\\
=&\frac{1}{\mathcal{A}}\sum_{\pp} e^{-i \pp \cdot (\RR'+\rr_0-\RR)} \jj(\pp).
\end{aligned}
\label{2center}
\end{equation}
where $\jj(\rr)$ is a function of displacement.  In replacing the first form of Eq.~(\ref{2center}) 
by the second we have assumed that the matrix element
depends only on difference between electron and hole positions.  The 
third form of Eq.~(\ref{2center}) simply expresses  $\jj(\rr)$ in terms of its Fourier transform.
The two-center approximation preserves all symmetries of the system, and
results derived based on this approximation are compatible with all symmetry requirements.
Using (\ref{ExWF}), (\ref{Bloch}) and (\ref{2center}), we can rewrite $\JJ$ in the form:
\begin{equation}
\begin{aligned}
\JJ(\theta, \rr_0,\QQ)=&\frac{N}{\mathcal{A}^2}\sum_{\kk}\sum_{\GG_B,\GG_T} 
f_{\kk}\\\times 
&\jj(\KK_B+\GG_B+\kk-m_h\QQ/M) \\
\times &\delta_{\QQ,(\KK_B+\GG_B)-(\KK_T+\GG_T)} e^{-i \GG_T\cdot \rr_0}.
\end{aligned}
\label{2centerkspace}
\end{equation}
Eq.~(\ref{2centerkspace}) establishes the momentum-conservation law in (\ref{conservation_law}).
Because of the relatively large 
separation between layers in the heterobilayer system
we expect that $\jj(\rr)$ should be a smooth function of $\rr$ on the scale of 
the monolayer lattice constant, and hence that 
$ \jj(\pp) $ should decline rapidly on the scale of the monolayer reciprocal lattice vector.

It follows from Eq.~(\ref{2centerkspace}) that the smallest excitation
momenta at which bright excitons appear  
are $\QQ=\qq_1$, $\qq_2$ and $\qq_3$, 
where $\qq_1=\KK_B-\KK_T$, and $\qq_{2}$ and $\qq_{3}$ are the three-fold-rotation 
counterparts of $\qq_1$ as illustrated in Fig.~\ref{Fig:Pattern}(e). 
The contributions to Eq.~(\ref{2centerkspace}) from larger momenta allowed by 
Eq.~(\ref{conservation_law}) will be much smaller because 
$\jj(\pp)$ is a decreasing function of $|\pp|$. 
Below we retain only the dominant optical response from
the $\qq_n$ contributions.  
This yields
\begin{equation}
\begin{aligned}
\JJ(\theta, \rr_0,\QQ) \approx \sum_{n=1}^{3}\delta_{\QQ,\qq_n}\JJ_n e^{-i \GG_T^{(n)}\cdot \rr_0},
\end{aligned}
\label{JJn}
\end{equation}
where $\GG_T^{(1)}$ is a zero vector, and $\GG_T^{(2, 3)}$  are reciprocal lattice vectors that connect equivalent $\KK$ points in the Brillouin zone of top layer ([Fig.~\ref{Fig:Pattern}(e)]). The parameters $\JJ_n$ 
are independent of $\rr_0$.

We extract numerical values for $\JJ_n$ by comparing with microscopic {\it ab initio} 
calculations performed for the commensurate structures obtained 
by setting $\theta$ and $\QQ$ to 0.  In this limit Eq.~(\ref{JJn}) reduces to 
\begin{equation}
\JJ(\rr_0)\approx \sum_{n=1}^{3} \JJ_n e^{-i \GG^{(n)}\cdot \rr_0}.
\label{JJrr0}
\end{equation}
In (\ref{JJrr0}), $\GG^{(n)}$ is $\lim_{\theta \to 0} \GG_T^{(n)}$.
Because the heterobilayer structure is commensurate in this limit, the {\it ab initio}
calculations are readily performed:
\begin{equation}
\begin{aligned}
\JJ(\rr_0)&=\frac{1}{\mathcal{A}}\sum_{\kk}f_{\kk} \langle G |\hat{\jj} a_{c (\KK+\kk)}^\dagger a_{v(\KK+\kk)}  | G \rangle \\
&\approx \tilde{f}(0) \DD(\rr_0),
\end{aligned}
\label{Jr0}
\end{equation}
where $\tilde{f}(0)=\sum_{\kk} f_{\kk}/\mathcal{A}$, and $\DD(\rr_0)=\langle v \KK | \boldsymbol{\mathcal{J}} | c \KK \rangle$.
$|\tilde{f}(0)|^2$ is the probability that the
electron and hole spatially overlap, and $\boldsymbol{\mathcal{J}}$ is the matrix 
representation of the current operator $\hat{\jj} $. 
In (\ref{Jr0}), the final expression is justified 
by the fact that $f_{\kk}$ is peaked 
around $\kk=0$. 
For simplicity, we assume $\tilde{f}(0)$ is independent of $\rr_0$, 
since the most significant dependence on $\rr_0$ is from $\DD(\rr_0)$ as 
we discuss below.  Not only the exciton energy but also the strength of 
optical coupling to the exciton state varies with position within the moir\'e pattern.


\section{Interlayer Optical Matrix Elements} 
\label{Sec:OME}
We performed {\it ab initio} calculations for aligned layers with
identical lattice constants as a function of $\rr_0$, which is the in-plane displacement shown in Fig.~\ref{Fig:Pattern}.  
The {\it ab initio} calculations are performed using fully-relativistic 
density-functional-theory (DFT) under the local-density approximation as implemented 
in \textsc{Quantum Espresso} \cite{giannozzi2009}; the details of these calculations are the 
same as those described in the Supplemental Material of Ref.~\cite{Wu2017}, except that we use a 
denser $18 \times 18 \times 1$ k-point sampling which is 
required to reduce noise in the individual conduction and valence band extrema in the AB 
stacking case. 
For each value of $\rr_0$, we fit the 
bands to a tight-binding model using Wannier90\cite{mostofi2014}, and then 
use the tight-binding model Hamiltonian to evaluate the matrix 
element of the current operator between the semiconductor ground state 
and the single exciton state using Eq.~(\ref{Jr0}).

\begin{figure}[t!]
	\includegraphics[width=1\columnwidth]{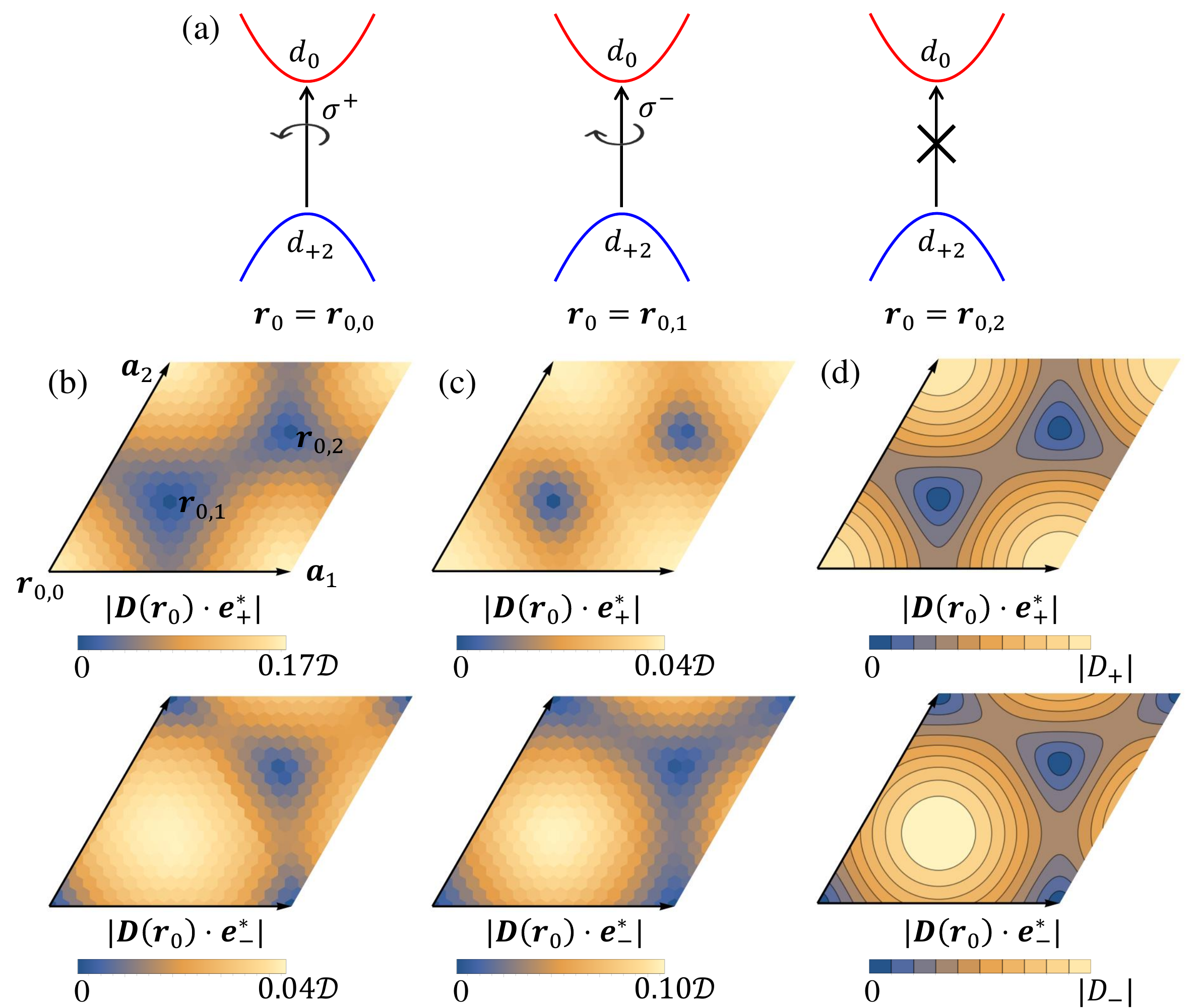}
	\caption{(a) $\KK$-valley interlayer exciton optical matrix elements 
	for heterobilayers with zero twist angle at special high symmetry 
	displacements.  (b) In-plane displacement dependence of the {\it ab initio} values 
	of $|\DD(\rr_0)\cdot \ee_+^*|$ (top) and $|\DD(\rr_0)\cdot \ee_-^*|$  (bottom)
	for aligned AA stacking of WS$_2$/MoS$_2$ with zero twist angle. 
	Values are given in units of the corresponding intralayer optical response parameter $\cal{D}$.
	$|\DD(\rr_0)\cdot \ee_-^*|$ at $\rr_{0,1}$ is small but nonzero.
	(c) Same plot as (b) but for AB stacking. 
	(d) Leading harmonic approximation for $|\DD(\rr_0)\cdot \ee_{\pm}^*|$ explained in the text.}
	\label{Fig:selection_rule}
\end{figure}

The bilayer has a $\hat{C}_3$ symmetry when $\rr_0$ has one of 
three high symmetry displacements: 
$\rr_{0,n}=n(\ba_1+\ba_2)/3$ for $n=0,1$ and 2.
Here $\ba_{1, 2}$ are the lattice vectors shown in Fig.~\ref{Fig:Pattern}(c).
At $\rr_{0,0}=0$, the metal atoms in the two layers are vertically aligned.
At $\rr_{0,1}$ ($\rr_{0,2}$) , Mo atoms in the top MoX$_2$ layer are above the chalcogen (empty) sites of the bottom WX$_2$ layer.
The $\hat{C}_3$ symmetry puts strong constraints on $\DD(\rr_{0,n})$. 
For both AA and AB stackings,
the main character of the valence (conduction) band 
at the $\KK$ valley is the W (Mo) $d$ orbital with magnetic quantum 
number +2 (0).
The optical selection rule is determined not only by the atomic magnetic 
quantum numbers, but also by the change in the Bloch phase factor 
under $\hat{C}_3$ operations. 
The Bloch phase shift is different for different $\rr_{0,n}$, leading
to different optical selection rules. 
Based on these considerations, we find that the transition at 
$\KK$ satisfies the following circular polarization selection rules:
\begin{equation}
\DD(\rr_{0,0})=D_+\ee_+, \DD(\rr_{0,1})=D_-\ee_-, \DD(\rr_{0,2})=0, 
\label{DDe}
\end{equation}
where $\ee_{\pm}=(1,\pm i)/\sqrt{2}$. Equation (\ref{DDe}) is valid for both AA and 
AB stacking. Note that valley-dependent optical selection rule is 
opposite for $\rr = \rr_{0,0}$ and  $\rr = \rr_{0,1}$, and that the in-plane transition dipole moment is zero for $\rr_0=\rr_{0,2}$. 
Fig.~\ref{Fig:selection_rule}(a) summarizes these optical selection rules 
schematically.

Using Eqs. (\ref{JJrr0})-(\ref{DDe}), we can
express $\JJ_n$ in the form:
\begin{equation}
\JJ_n=\frac{1}{3}[J_+\ee_+ + e^{i \GG^{(n)}\cdot \rr_{0,1}} J_-\ee_-],
\label{Jn}
\end{equation}
where $J_\pm= \tilde{f}(0) D_\pm$. Because $\JJ_n$ has both $\ee_+$ and $\ee_-$ components, it 
follows that the optically active 
exciton states at excitation wavevector $\qq_n$ have
elliptical optical selection rules, instead of circular \cite{Yu2015}. 
However, as we show below, the exciton potential energy produced by the moir\'e pattern will  
mix the three states located at $\qq_n$ ($n=$ 1, 2, 3). 
The resulting eigenstates 
states are coherent superposition of the $|\qq_n \rangle$ states for 
which circular optical selection rules are restored.
Substituting Eq.~(\ref{Jn}) back into Eq.~(\ref{JJrr0}), we can 
parametrize $\DD(\rr_0)$ in terms of $D_\pm$:
\begin{equation}
\begin{aligned}
\DD(\rr_0)\approx  & \Big[ \frac{D_+}{3}\sum_{n=1}^{3} e^{-i \GG^{(n)}\cdot \rr_0} \Big] \ee_+ \\
+& \Big[ \frac{D_-}{3}\sum_{n=1}^{3} e^{-i \GG^{(n)}\cdot (\rr_0-\rr_{0,1})} \Big] \ee_-.
\end{aligned}
\label{DDrr0}
\end{equation}

Equation~(\ref{DDrr0}) can be tested numerically. The DFT values of $|\DD(\rr_0)\cdot \ee^*_{\pm}|$ are presented in 
Figs.~\ref{Fig:selection_rule}(b) and \ref{Fig:selection_rule}(c), respectively for AA and AB 
stacking of the WS$_2$/MoS$_2$ bilayer. In Fig.~\ref{Fig:selection_rule}(d) 
we also plot $|\DD(\rr_0)\cdot \ee^*_{\pm}|$, as calculated from 
Eq.~(\ref{DDrr0}).
The good agreement between analytical and numerical results 
provides strong justification for Eq.~(\ref{DDrr0}).
From the {\it ab initio} results we estimate that $(|D_+|, |D_-|)$ is $(0.17,0.04)\mathcal{D}$ for AA stacking and $(0.04,0.10)\mathcal{D}$ for AB stacking, where $\mathcal{D}$ is the matrix element for MoS$_2$ intralayer excitations
in AA stacking with zero twist angle and zero displacement.  
$\mathcal{D}$ is defined as $|\langle v \KK |\boldsymbol{\mathcal{J}}| c \KK \rangle |$, where both the 
conduction and valence states are located in MoS$_2$ layer. 
The numerical value of $\hbar \mathcal{D}/e$ is about 5.22 eV$\cdot$\AA, where $\hbar$ is the 
reduced Planck constant and $e$ is the electron charge.
Numerical results of WSe$_2$/MoSe$_2$ bilayer are presented in the Appendix \ref{appC}.

\section{Interlayer Exciton Potential Energy} 
\label{Sec:Potential}
\begin{figure}[t]
	\includegraphics[width=1\columnwidth]{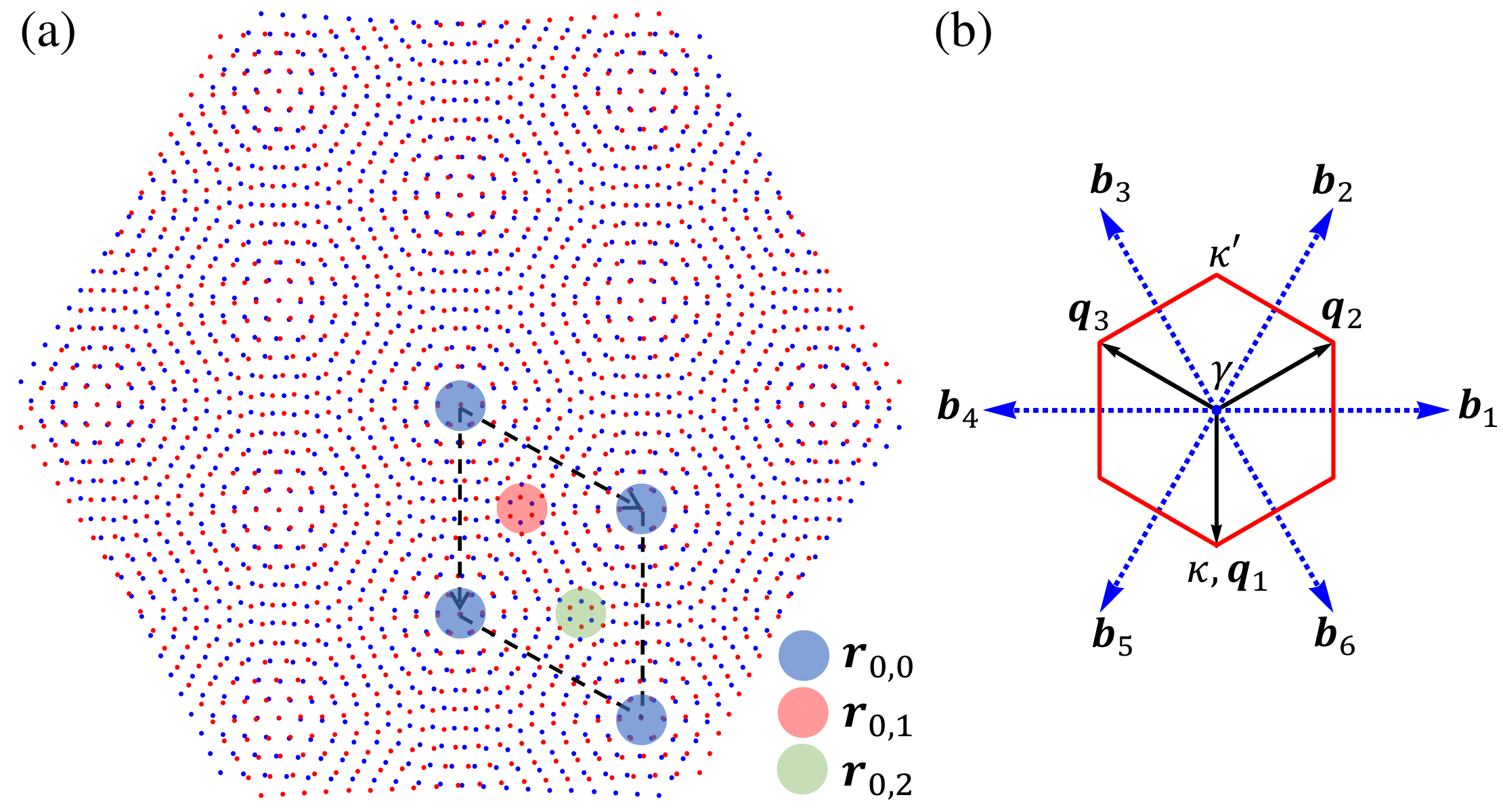}
	\caption{(a) The small red and blue dots represent Bravais lattices of the two layers with a relative rotation. 
	             The dashed lines mark the moir\'e unit cell.
	             The big dots indicate regions where the local displacement between the two layers is $\rr_{0,n}$.
	             Each $\rr_{0,1}$ region has three $\rr_{0,2}$ neighbors and each 
	              $\rr_{0,2}$ region has three $\rr_{0,1}$ neighbors.
	         (b) Moir\'e Brillouin zone and the first-shell moir\'e reciprocal lattice vectors.}
	\label{Fig:Lattice_BZ}
\end{figure}

\begin{figure*}[t!]
	\includegraphics[width=1.8\columnwidth]{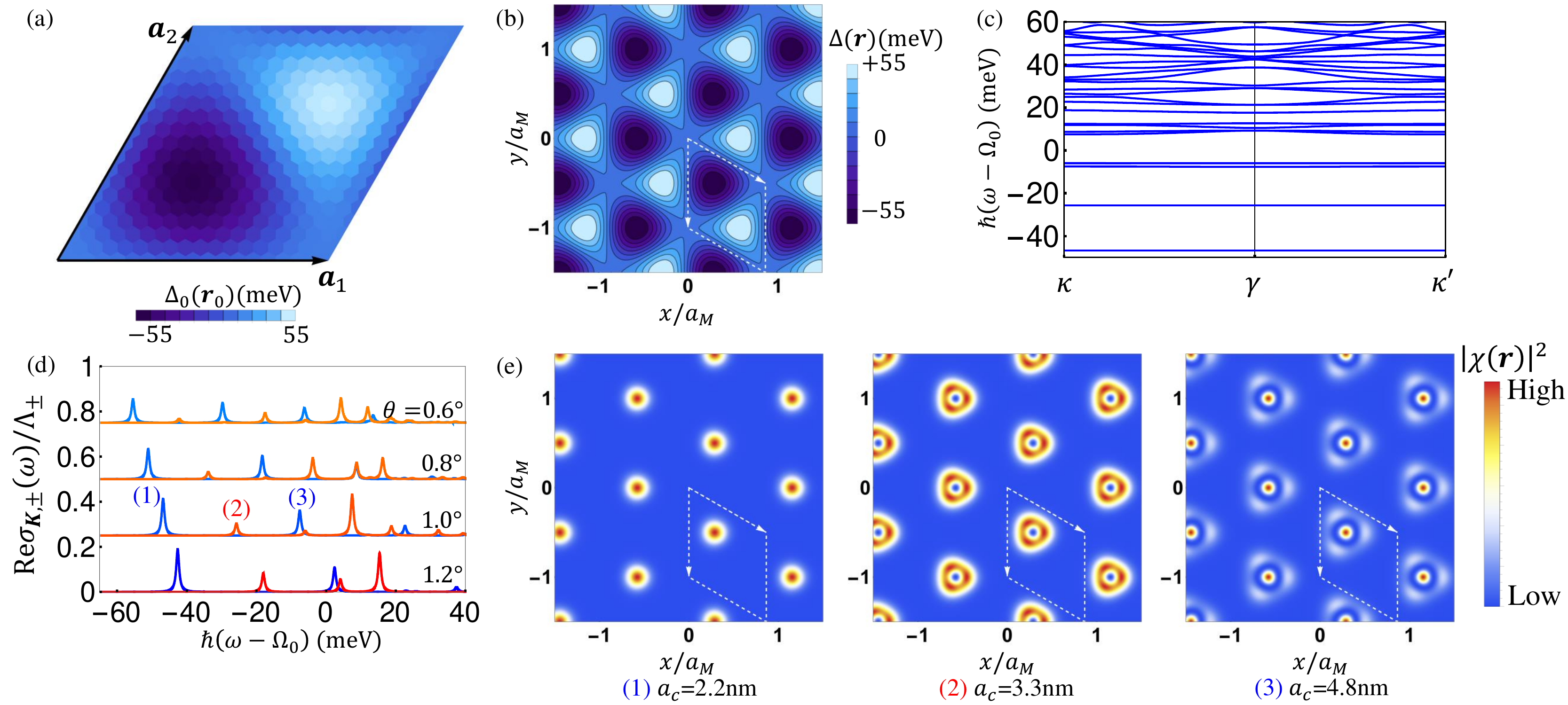}
	\caption{Results for a WS$_2$/MoS$_2$ heterobilayer with AA stacking.  
	(a) Variation of the interlayer band gap as a function of $\rr_0$ when the twist angle is zero. (b) Variation of the band gap in real space when the twist angle is finite. (c) Exciton moir\'e band structure for $\theta=1^{\circ}$. (d) Optical spectrum of $\KK$-valley interlayer excitons for a series of twist angles. The red and blue curves respectively show $\text{Re}\sigma_{\KK,+}/\Lambda_+$ and $\text{Re}\sigma_{\KK,-}/\Lambda_-$. $\Lambda_+/\Lambda_-$ is about 18. The broadening parameter $\eta$ is taken to be 0.5 meV. Curves for different $\theta$ are shifted vertically.
 (e) The real space probability function $|\chi(\rr)|^2$ of the interlayer excitons responsible for the first three absorption peaks at $\theta=1^{\circ}$.}
	\label{Fig:AA_spectrum}
\end{figure*}

The interlayer twist generates not only a relative shift in momentum space[Fig.~\ref{Fig:Pattern}(e)],
but also a moir\'e pattern in real space [Fig.~\ref{Fig:Lattice_BZ}(a)]. 
The moir\'e pattern leads to spatial modulation of the exciton energy.
The {\em potential} energy for the interlayer excitons is $E_g-E_b$, where $E_g$ is the band gap between the WX$_2$ valence band and the MoX$_2$ conduction band at the $\KK$ point, 
and $E_b$ is the exciton binding energy. We neglect the variation of $E_b$ in the moir\'e pattern, since it is 
typically smaller than the variation of $E_g$. We use a local 
approximation\cite{Jung2014,wu2014tunable, Wu2017} to estimate the exciton moir\'e potential. 
First we obtain $E_g$ as a function of relative displacement $\rr_0$ at {\it zero} twist angle from the 
DFT band structure. 
The spatial variation of $E_g$ is shown in Figs.~\ref{Fig:AA_spectrum}(a) and
 \ref{Fig:AB_spectrum}(a).  Note that the exciton potential energy has its minimum at the $\rr_{0,1}$ point for
 AA stacking and at the $\rr_{0,0}$ point for AB stacking.   
Since the potential is a smooth periodic function of $\rr_0$ in both cases, we can 
approximate it by the lowest order harmonic 
expansion:
\begin{equation}
\begin{aligned}
\Delta_0(\rr_0)    \equiv E_g(\rr_0)- \langle {E_g} \rangle
 \approx \sum_{j=1}^6 V_j \exp( i \bold{G}_j\cdot \rr_0),
\end{aligned}
\end{equation}
where $\langle {E_g} \rangle$ is the average of $E_g$ over $\rr_0$, and the $\bold{G}_j$ 
are the first-shell reciprocal lattice vectors shown in Fig.~\ref{Fig:Lattice_BZ}(c). 
Because each layer {\it separately} has $\hat{C}_3$ symmetry, 
$\Delta_0$ is invariant under 120$^{\circ}$ rotations of $\rr_0$.
This symmetry implies that:
\begin{equation}
V_1=V_3=V_5, V_2=V_4=V_6.
\end{equation}
Because $\Delta_0$ is always real, $V_1=V_4^*$ is also required. 
As a result, all six $V_j$ are determined by $V_1=V\exp(i \psi)$.
We find that for MoS$_2$/WS$_2$ heterobilayers
$(V, \psi)$ is (12.4meV, 81.5$^{\circ}$) for AA stacking and (1.8meV, 154.5$^{\circ}$) for AB stacking.
The variation of the band gap predicted by DFT has been found to agree
reasonably well with scanning tunneling microscopy measurement.\cite{Zhang_Shih}

When the two layers have a relative twist angle $\theta$, the lattice sites associated with top and bottom layers are related by: $\RR_T= \mathcal{R}(\theta) \RR_B+\rr_0$, where  $\mathcal{R}(\theta)$ is the rotation matrix. The local displacement of the top layer relative to the bottom layer near position $\RR_T$ is given by
\begin{equation}
\begin{aligned}
\dd(\RR_T)&=\hat{T}_B (\RR_T)=\hat{T}_B (\RR_T-\RR_B) \\
& \approx  \hat{T}_B [\theta \hat{z}\times (\RR_T-\rr_0)+\rr_0],
\end{aligned} 
\label{ddRRT}
\end{equation}
where the operator $\hat{T}_B$ reduces a vector to the Wigner-Seitz
cell of the $\RR_B$ lattice. When $\theta$ is small, the displacement varies smoothly in real space and gives rise to the moir\'e pattern.
We assume that the variation of the band gap locally follows the variation of the displacement:
\begin{equation}
\begin{aligned}
\Delta(\rr)&=\Delta_0[\dd(\rr)] \approx \sum_{j=1}^6 V_j \exp[ i \bold{G}_{B,j}\cdot \dd(\rr)]\\
&\approx \sum_{j=1}^6 V_j \exp[i \bb_j \cdot (\rr+\rr_0 \times \hat{z} /\theta-\rr_0/2)],
\end{aligned}
\label{Deltarr}
\end{equation}
where $\bold{G}_{B,j}$ is the first-shell reciprocal lattice vectors associated with the bottom layer.
$\bb_j=\theta\bold{G}_{j}\times\hat{z}$ is the reciprocal lattice vector of the moir\'e pattern, 
illustrated in Fig.~\ref{Fig:Lattice_BZ}(b). 
The moir\'e periodicity is inversely proportional to the twist angle: $a_M\approx a_0/\theta$.
The last expression in (\ref{Deltarr}) implies that the global displacement $\rr_0$ just leads to a spatial translation of the $\Delta(\rr)$ potential when $\theta$ is finite. Without loss of generality, we take $\rr_0$ to be zero for the twisted bilayer in the following.

Including both kinetic and potential energy contributions 
the exciton Hamiltonian is 
\begin{equation}
\begin{aligned}
&H=\hbar\Omega_0+\frac{\hbar^2\QQ^2}{2M}+\Delta(\rr),\\
\end{aligned}
\label{moireham}
\end{equation}
where $\hbar\Omega_0$ is a constant, $\hbar^2\QQ^2/(2M)$ is the exciton 
center-of-mass kinetic energy.
The periodic potential $\Delta(\rr)$ is plotted in Figs.~\ref{Fig:AA_spectrum}(b) and \ref{Fig:AB_spectrum}(b). 
We have numerically diagonalized this Hamiltonian using a plane-wave expansion. 
The exciton band structure in the MBZ is shown in 
Figs.~\ref{Fig:AA_spectrum}(c) and \ref{Fig:AB_spectrum}(c) respectively for 
AA and AB stackings with a 1$^{\circ}$ twist angle.
The moir\'e potential mixes excitons with different momenta, leading to a dramatic change in the 
interlayer absorption spectrum as we explain in the next section.

\section{Interlayer Exciton Optical Absorption Spectrum}
\label{Sec:Oabsorb}
To study the effects of the moir\'e potential on the optical response,
it is  instructive to first consider the weak potential limit and use perturbation theory. 
As shown in Eq. (\ref{JJn}), there are three optically active exciton states, which are at momentum $\qq_n$, {\it i.e.}
at the corners of the MBZ [Fig~\ref{Fig:Lattice_BZ}(b)]. 
These three states have the same kinetic energy, and would be energetically 
degenerate if there were no moir\'e potential energy. 
However, the three states are coupled by the moir\'e potential because their momenta differ by moir\'e reciprocal lattice vectors. Therefore, the potential lifts the degeneracy. In degenerate perturbation theory, the potential projected to 
the $|\qq_n\rangle$ has the matrix form:
\begin{equation}
\begin{aligned}
\mathcal{V}=&\begin{pmatrix}
0    & V_5 & V_6\\
V_2  &   0 & V_1\\
V_3  & V_4 & 0  
\end{pmatrix}.
\end{aligned}
\end{equation}
The eigenstates of $\mathcal{V}$ have the form:
\begin{equation}
|\Phi_\lambda\rangle=\frac{1}{\sqrt{3}}(|\qq_1\rangle+e^{i 2\lambda \pi/3}|\qq_2\rangle+e^{-i 2\lambda \pi/3}|\qq_3\rangle),
\end{equation}
where $\lambda$ takes the values 0 and $\pm1$. 
The corresponding eigenvalue is $2V\cos(\psi+2\lambda\pi/3)$. As a result, different $|\Phi_\lambda\rangle$ generally have different energies unless $\psi$ is fine tuned to multiples of $\pi/3$. The optical matrix element for  $|\Phi_\lambda\rangle$ is given by:
\begin{equation}
\frac{1}{\sqrt{\mathcal{A}}}\langle G |\hat{\jj} | \Phi_\lambda \rangle = \frac{1}{\sqrt{3}} (\JJ_1+e^{i 2\lambda \pi/3}\JJ_2+e^{-i 2\lambda \pi/3}\JJ_3),
\label{PhiL}
\end{equation}
which is  $J_+\ee_+/\sqrt{3}$ for $\lambda=0$,
$J_-\ee_-/\sqrt{3}$ for $\lambda=+1$,
and vanishes  for $\lambda=-1$.
Since we have set the global displacement $\rr_0$ to be zero,
the  $\JJ_n$ in (\ref{PhiL}) are the optical matrix elements for $|\qq_n\rangle$.
We see from Eq. (\ref{PhiL}) that circular optical selection rules are 
restored by scattering off the moir\'e potential.
In particular, the exciton states $| \Phi_0 \rangle$ and $| \Phi_{+1} \rangle$ couple respectively to 
$\sigma_+$ and $\sigma_-$ circularly polarized light, while $| \Phi_{-1} \rangle$ is optically dark.
When $\psi$ is fine tuned to $2\pi/3$, $| \Phi_0 \rangle$ and $| \Phi_{+1} \rangle$ becomes energetically 
degenerate, which gives rise to Dirac cone in the exciton band structure near $\kappa$ point.
We will discuss this special situation in the next section.

\begin{figure*}[t!]
	\includegraphics[width=1.8\columnwidth]{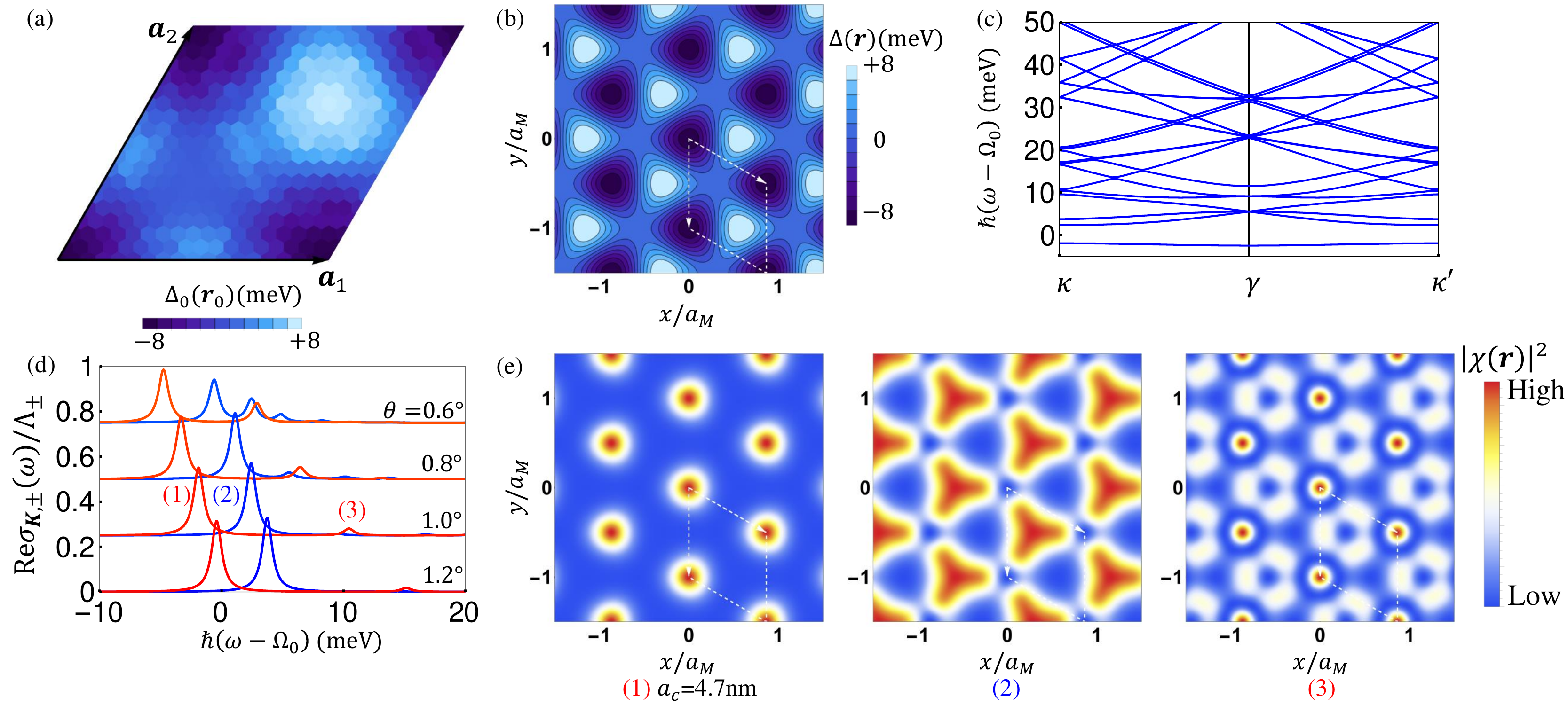}
	\caption{Same plots as those in Fig.~\ref{Fig:AA_spectrum} but for the 
	 WS$_2$/MoS$_2$ heterobilayer in AB stacking. 
	 In (d), $\Lambda_+/\Lambda_-$ is about 6.}
	\label{Fig:AB_spectrum}
\end{figure*}

The above analysis assumes that the potential is weak. We now 
argue that circular optical selection rules are generally expected for the twisted bilayer. 
When $\rr_0=0$, the bilayer with any twist angle $\theta$ has a $\hat{C}_3$ symmetry. 
Exciton states can be classified by the  symmetry: $\hat{C}_3 |\chi\rangle = \exp[i 2\pi l_\chi/3]|\chi\rangle$, where integer $\l_\chi$ is 0 or $\pm 1$.
Due to the  $\hat{C}_3$ symmetry, the excitation from ground state to exciton $|\chi\rangle$ is coupled to $\sigma_\pm$ circularly polarized light if $l_\chi=\pm 1$, and optically forbidden under normal incident light if $l_\chi=0$. 
In Appendix~\ref{appB} we explicitly show that the finite $\rr_0$  and 
zero $\rr_0$ problems are related by a unitary transformation.

Our {\it ab initio} results indicate that the amplitude of the moir\'e potential can be as large as 110meV, which is not weak at all.
Due to the moir\'e potential, the $|\qq_n\rangle$ are redistributed to different exciton states $|\chi_\alpha\rangle$ at the $\kappa$ point.
We assume that only the  $|\qq_n\rangle$  components in  $|\chi_\alpha\rangle$ contribute to the optical response. The optical conductivity can then be expressed as follows:
\begin{equation}
\begin{aligned}
&\text{Re}\sigma_{\KK,\pm}(\omega)\\ = &  \frac{1}{\omega \mathcal{A}}\sum_{\alpha} \Big |\langle \chi_\alpha  | \jj \cdot \ee_\pm| G \rangle \Big|^2\Gamma_1(\omega-\omega_\alpha)\\
 \approx & \frac{1}{\omega}\sum_{\alpha} \Big | \sum_{n=1}^3 \langle \qq_n |\chi_\alpha \rangle \JJ_n \cdot \ee_\pm^* \Big|^2 \Gamma_1(\omega-\omega_\alpha)\\
\approx & \Lambda_\pm \sum_{\alpha} \Big | \frac{1}{3}\sum_{n=1}^3 e^{ i \phi_{\pm}^{(n)}} \langle \qq_n |\chi_\alpha \rangle  \Big|^2 \Gamma_2(\omega-\omega_\alpha) 
\end{aligned}
\label{sigma}
\end{equation}
where $\Gamma_m(\omega-\omega_\alpha)=\eta^m/\big[\hbar^2(\omega-\omega_\alpha)^2+\eta^2\big]$, $\hbar \omega_\alpha$ is the energy of state $|\chi_\alpha\rangle$,
and $\eta$ is a broadening parameter. In (\ref{sigma}), $\Lambda_{\pm}=|J_\pm|^2/(\eta \Omega_0)$, and the phases are respectively $\phi_{+}^{(n)}=0$ and $\phi_{-}^{(n)}=\GG^{(n)}\cdot\rr_{0,1}$.
$\text{Re}\sigma_{\KK,\pm}(\omega)$ measures the optical absorption by $\KK$-valley excitons in response to $\sigma_\pm$ polarized light at frequency $\omega$.
The optical response from $-\KK$ valley can be obtained by time-reversal symmetry: $\text{Re}\sigma_{-\KK,\mp}=\text{Re}\sigma_{\KK,\pm}$.

Theoretical optical conductivities calculated for small twist angles are shown in 
Figs.~\ref{Fig:AA_spectrum}(d) and 
\ref{Fig:AB_spectrum}(d). The optical spectrum has the following features. (i) both $\text{Re}\sigma_{\KK,+}$ and $\text{Re}\sigma_{\KK,-}$ have peaks around energy $\hbar\Omega_0$, but their peaks are located at distinct energies. This is a manifestation of the circular optical selection rule. For example, the $\KK$-valley exciton state in peak (1) of Fig.~\ref{Fig:AA_spectrum}(d) can be excited by $\sigma_-$ polarized light, but not by $\sigma_+$ polarized light. (ii) There are a series of peaks instead of a single peak in $\text{Re}\sigma_{\KK,+}$ ($\text{Re}\sigma_{\KK,-}$) near energy $\hbar\Omega_0$. Umklapp scattering off the moir\'e potential makes finite-momentum excitons optically 
active. (iii) The optical spectrum has a strong dependence on twist angle because it 
tunes the moir\'e periodicity in real space and the MBZ in momentum space, 
leading to changes in the exciton moir\'e band structure and optical response.

To gain a deeper insight, we study the exciton center-of-mass wave function $\chi_\alpha(\rr)$, 
which is defined in real space as follows:
\begin{equation}
\chi_\alpha(\rr)=\sum_{\QQ} \langle \QQ | \chi_\alpha \rangle e^{i \QQ \cdot \rr}.
\end{equation}
Fig.~\ref{Fig:AA_spectrum}(e) [\ref{Fig:AB_spectrum}(e)] plots  $|\chi_\alpha(\rr)|^2$ for excitons in the first three peaks
of Fig.~\ref{Fig:AA_spectrum}(d) [\ref{Fig:AB_spectrum}(d)].      
The amplitude of the moir\'e potential is about 110meV for AA stacking. In such a strong potential, the exciton wave function is composed of Wannier states that are strongly confined to the potential minimum positions, as shown in Fig.~\ref{Fig:AA_spectrum}(e).
The energy dispersion of the first few exciton bands is almost flat [Fig.~\ref{Fig:AA_spectrum}(c)], indicating that tunneling between neighboring Wannier states is negligibly weak.
For AB stacking the moir\'e potential is weaker, but a similar confinement is found for  
excitons in peak (1); see Fig.~\ref{Fig:AB_spectrum}(e).

To characterize the localization length of low-energy excitons, we define a parameter $a_c$ as:
\begin{equation}
a_c=\sqrt{\int d\rr |\delta \rr|^2 |\chi_\alpha(\rr)|^2 / \int d\rr |\chi_\alpha(\rr)|^2},
\end{equation}
where the spatial integral is restricted to one moir\'e Wigner-Seitz cell centered at a potential minimum position, 
and $|\delta \rr|$ is the distance relative to the center.  
For the exciton state in peak (1) of Fig.~\ref{Fig:AA_spectrum}(d), $a_c$ is 2.8nm at $\theta=0.6^{\circ}$, and decreases  to 2.3nm at $\theta=1.0^{\circ}$. For comparison, the moir\'e periodicity $a_M$ is about 30.5nm and 18.3nm respectively for $\theta=0.6^{\circ}$ and $1.0^{\circ}$. Values of $a_c$ for other exciton states are listed in Figs.~\ref{Fig:AA_spectrum}(e) and \ref{Fig:AB_spectrum}(e).
Here $a_c$ of the localized exciton states is not significantly larger than exciton internal radius $a_X$ ( $a_X$ is about 1.3 nm as estimated in Appendix~\ref{appA}). The relative motion between electron and hole will be affected by the moir\'e potential. This effect is neglected in our theory. A more accurate approach is to solve the exciton problem for a supercell structure.
Such a theory will still lead to localized exciton states due to the
moir\'e potential, so we expect that the qualitative picture of our theory remains valid.

The localized Wannier states can be understood using a model of particles in a two-dimensional parabolic potential.
Near its minima, the moir\'e potentials can be approximated as parabolic: 
$\beta (\delta \rr/a_M)^2/2$, where $\beta$ is a constant independent of $a_M$. A particle with mass $M$ in the parabolic potential has energies $\sqrt{\beta \hbar^2/(M a_M^2) }(n_x+n_y+1)$ , where $n_{x,y}$ are non-negative integers.  The quantized energy levels explain why peaks (1), (2) and (3) in Fig.~\ref{Fig:AA_spectrum}(e) have nearly equal energy spacing. Wave function  $\Psi_{00}$ of the lowest-energy level is proportional to $\exp[-\delta \rr^2/(2\ell^2)]$,
where $\ell=[\hbar^2 a_M^2/ (\beta M)]^{1/4}$. $\Psi_{00}$ has an $s$-wave symmetry and provides a good approximation for the exciton in peak (1) of both AA and AB stackings. The localization length $a_c$ 
estimated based on $\Psi_{00}$ is equal to $\ell$, which captures the 
dependence of $a_c$ on the moir\'e periodicity $a_M$.
The exciton in peak (2) of AA stacking corresponds to a chiral $p$-wave 
state of the first excited level in the parabolic potential,
while the exciton in peak (3) mimics a zero-angular-momentum 
state associated with the second excited level; see Fig~\ref{Fig:AA_spectrum}(e). 

\begin{figure*}[t!]
	\includegraphics[width=1.8\columnwidth]{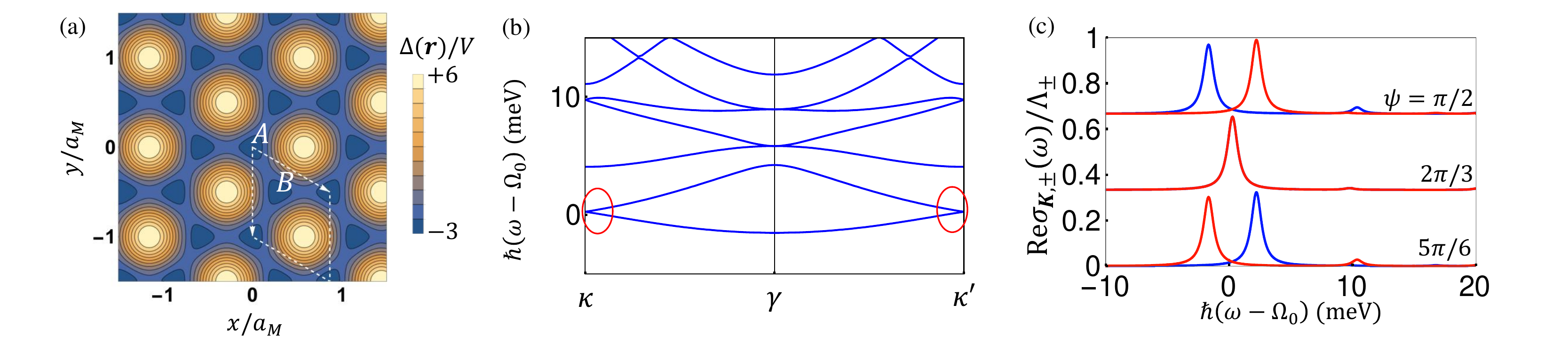}
	\caption{(a) Exciton potential with $\psi=2\pi/3$. $A$ and $B$ are potential minimum positions that form a honeycomb lattice.
	(b)Exciton band structure when $\psi=2\pi/3$. The red circles highlight the Dirac cones.
	(c)Optical spectrum for $\psi$ below, at and above $2\pi/3$. The red and blue curves respectively show $\text{Re}\sigma_{\KK,+}/\Lambda_+$ and $\text{Re}\sigma_{\KK,-}/\Lambda_-$. Curves for different $\psi$ are shifted vertically.
 In (b) and (c), the twist angle is $1^{\circ}$. Other parameters take their values in AB stacked WS$_2$/MoS$_2$ bilayer.}
	\label{Fig:Dirac_cone}
\end{figure*}

Finally, we reinterpret the optical response in terms of the real space wave function. 
The optical matrix element of state $|\chi_\alpha\rangle$ is:
\begin{equation}
\begin{aligned}
\frac{1}{\sqrt{\mathcal{A}}}\langle G | \hat{\jj} | \chi_\alpha \rangle
=&\sum_{n=1}^{3} \JJ_n \langle \qq_n |\chi_\alpha \rangle\\
=&\frac{1}{\mathcal{A}}\int d\rr \Big( \sum_{n=1}^{3} \JJ_n e^{- i \qq_n \cdot \rr}\Big) \chi_\alpha(\rr)\\
=&\frac{1}{\mathcal{A}}\int d\rr \tilde{\JJ}(\rr) u_\alpha(\rr),
\end{aligned}
\label{spatialaverage}
\end{equation} 
where $\tilde{\JJ}(\rr)$ represents $\JJ_1+\JJ_2 \exp(-i \bb_2 \cdot \rr)+\JJ_3 \exp(-i \bb_3 \cdot \rr)$, and $u_\alpha(\rr)$, the periodic part of $\chi_\alpha$, is equal to $\exp(-i \qq_1 \cdot \rr) \chi_{\alpha}(\rr)$.
Both the magnitude and polarization of  $\tilde{\JJ}(\rr)$ change in real space. The variation follows that of the local displacement, because $\tilde{\JJ}(\rr)$ is equivalent to $\JJ[\dd(\rr)]$, where 
functions $\JJ(\rr_0)$ and $\dd(\rr)$ are respectively defined in (\ref{JJrr0}) and (\ref{ddRRT}).
 $\tilde{\JJ}(\rr)$ generally has both $\ee_+$ and $\ee_-$ components, but becomes fully circularly polarized at high-symmetry positions where the local displacement $\dd(\rr)$ is $\rr_{0,0}$ or $\rr_{0,1}$. 
 
Equation (\ref{spatialaverage}) indicates that the optical response is an average of $\tilde{\JJ}$ over space weighted by the exciton center-of-mass wave function $u_\alpha(\rr)$. Different forms of $u_\alpha(\rr)$ give rise to distinct responses. 
For example, excitons in both peak (1) and (2) [Fig.~\ref{Fig:AA_spectrum}(d)] are localized around the potential minimum positions. 
However, they couple to light with different circular polarization, 
because their localized Wannier states have respectively $s$ and $p$ wave symmetries. 
Equation (\ref{spatialaverage}) can be generalized to the case where a heterobilayer has a finite system size.

\section{Dirac cones in exciton band structure}
\label{Sec:Dirac}
As mentioned above, there is a degeneracy between the two lowest energy states at $\kappa$ point when $\psi$ is fine tuned to $2\pi/3$. At $\psi=2\pi/3$, the exciton potential $\Delta(\rr)$ has an additional two-fold rotational symmetry besides the $\hat{C}_3$ symmetry, and creates an effective honeycomb lattice for excitons, as illustrated in Fig.~\ref{Fig:Dirac_cone}(a). 
It is well known from study of graphene that a honeycomb lattice harbors Dirac cones in band structure.
By explicit calculation using our continuum model, we do find Dirac cones located around $\kappa$ and $\kappa'$ point; see Fig.~\ref{Fig:Dirac_cone}(b).

The two sublattices of the honeycomb lattice are located at positions $A$ and $B$ [Fig.~\ref{Fig:Dirac_cone}(a)].
When $\psi=2\pi/3$, $A$ and $B$ positions are related by the two-fold rotational symmetry, and thus have the same potential energy $\Delta_A=\Delta_B$. On the other hand, the local optical selection rules are opposite for $A$ and $B$; see Eq.~(\ref{spatialaverage}). 
Because of the additional  two-fold rotational symmetry, $\text{Re}\sigma_{\KK,+}/\Lambda_+$  and $\text{Re}\sigma_{\KK,-}/\Lambda_-$ have energetically degenerate response peaks, as shown in Fig.~\ref{Fig:Dirac_cone}(c).

When $\psi$ is away from $2\pi/3$, the two-fold rotational symmetry is broken, leading to difference between $\Delta_A$ and $\Delta_B$.
The staggered potential $(\Delta_{AB}=\Delta_A-\Delta_B)$ gaps out the Dirac cone.
The sign of $\Delta_{AB}$ is controlled by the parameter $\psi$.
For $\psi$ slightly below (above) $2\pi/3$, $\Delta_{AB}$ is positive (negative).
The optical spectrum can have a strong variation when $\psi$ crosses $2\pi/3$, as indicated in Fig.~\ref{Fig:Dirac_cone}(c).

We emphasize again  that optically active excitons of valley $\KK$ ($-\KK$) are located at $\kappa$ ($\kappa'$) point in our convention.
Therefore, the Dirac cones, if present, are directly relevant to optical responses.
The Dirac cones are generally gapped out, which can still lead to interesting effects.
For example, a domain wall, which separates two domains with opposite potential difference $\Delta_{AB}$, will host excitonic valley-momentum locked helical states. We leave a detailed study of the helical state to future work.

\section{Discussion}
\label{Sec:Disc}
Our theory applies to small twist angle in the range between $0.5^{\circ}$ and $2^{\circ}$, where moir\'e pattern has a strong influence on electronic and optical properties. The twist angle can be controlled to $0.1^{\circ}$ accuracy in the case of bilayer graphene using a technique developed in Ref.~\onlinecite{Kim_bilayer}. Similar controlled technique could be generalized to TMD bilayers.
We hope our work can stimulate systematic experimental study of TMD bilayers with small twist angle.  

We briefly discuss some experimental implications of our work. 
One of our important findings is that interlayer excitons have optical selection rules that are not locked to valleys, as illustrated in Figs.~\ref{Fig:AA_spectrum}(d) and \ref{Fig:AB_spectrum}(d). 
To demonstrate the experimental consequences, we consider AA stacking for definiteness. 
In AA stacking, the {\it intralayer} $A$ excitons of both MoX$_2$ and WX$_2$ layers at the $\KK$ valley are coupled to 
$\sigma_+$ circularly polarized light. In a photoluminescence experiment, $\sigma_+$ circularly polarized light in resonance with MoS$_2$ $A$ exciton will excite the intralayer excitons at the $\KK$ valley.  
The $A$ exciton will then relax to form interlayer excitons through hole transfer across the 
MoS$_2$/WS$_2$ bilayers. In the transfer process, the carrier spin and valley indices are expected to be 
conserved.\cite{Schaibley2016} Therefore, interlayer excitons are formed predominantly in valley $\KK$. 
Reversed circular polarization emission occurs for certain interlayer excitons. 
For example, the interlayer exciton in peak (1) of Fig.~\ref{Fig:AA_spectrum}(d) will emit $\sigma_-$ circularly polarized light, which is opposite to the polarization of the excitation light.  The physics  
of the processes that control the steady state probability 
distribution of interlayer excitons under continuous intralayer excitation is clearly
extremely rich, and this will ultimately
control the polarization distribution of the outgoing light.
Because many of the moir\'e bands 
have minima away from the optically active moir\'e Brillouin-zone corners,
the moir\'e pattern may cause more excitons to accumulate in optically 
dark states.  

The moir\'e potential can lead to localized exciton states that are confined to the potential minimum positions, as illustrated in Figs.~\ref{Fig:AA_spectrum}(e) and \ref{Fig:AB_spectrum}(e). This opens the door to scalable engineering of a two-dimensional array of quantum dots using twisted TMD heterobilayers. The effective radius of the quantum dot can be estimated in terms of the parameter $a_c$, while the inter-dot separation is given by $a_M$. Both $a_c$ and $a_M$ are tunable by twist angle. As discussed in Sec.~\ref{Sec:Oabsorb}, $a_c$ and $a_M$ are respectively about 3 nm and 30 nm for AA stacking with $\theta=0.6^{\circ}$. 
The nanoscale quantum dots could be studied using single molecule
localization microscopy, which has a nanometer resolution\cite{JFeng2017}.
When fine tuned, the moir\'e potential gives rise to Dirac excitons, which could be utilized to design one-dimensional excitonic channels.
The twisted heterobilayers provide a new platform to study excitons, polaritons, and their condensate in a triangular lattice,
where Bose Hubbard model physics can be explored.\cite{Bose_Hubbard} 

Scanning tunneling microscopy measurement \cite{Zhang_Shih} has identified a spatial modulation with an amplitude of 150meV in the local band gap of a MoS$_2$/WSe$_2$ heterobilayer with rotational alignment and lattice mismatch.
Our theory can be generalized to this case, and similar properties of interlayer excitons are expected.

In conclusion, optical absorption by interlayer excitons in TMDs is split into subfeatures with both senses of circular selectivity in a given valley.  The overall spectrum and its optical selectivity are both sensitive to spatial variation of  the exciton potential energy and the local absorption strength.  Photoluminescence and absorption studies of these systems are expected to provide a rich characterization of twisted heterobilayers.

\section{Acknowledgment}
F. W. thanks Xiaoqin Li for valuable discussions. 
The work of F. W. at Argonne was supported by Department of Energy, Office
of Basic Energy Science, Materials Science and Engineering Division.
Work at Austin was supported by the US Army Research Office under MURI award W911NF-17-1-0312,
and by the Welch Foundation under Grant No. TBF1473.
The authors acknowledge the Texas Advanced Computing Center (TACC) at The University of Texas at Austin for providing HPC resources that have contributed to the research results reported within this paper.

{\it Note added}.
When finalizing the manuscript, we became aware of a related work by Yu {\it et al}.\cite{Yu2017}
While a different approach was taken in that paper, our results appear to agree where they overlap.
We have explicitly worked out the optical absorption spectrum in this manuscript.

\appendix
\section{Electron-hole relative motion wave function}
\label{appA}
Electron and hole in one exciton are bound by the attractive Coulomb interaction.
In a parabolic band approximation, which is accurate when the binding energy is small compared to the band gap, the electron-hole relative motion wave function $f(\kk)$ is determined by:
\begin{equation}
\sum_{\kk'}\Big[\frac{\hbar^2 \kk^2}{2\mu}\delta_{\kk \kk'}
-\frac{1}{\mathcal{A}}U(\kk-\kk')\Big]f(\kk')=-E_b f(\kk),
\label{Xwave}
\end{equation}
where $\mu=m_e m_h/(m_e+m_h)$ is the reduced mass and $E_b$ is the electron-hole binding energy. 
In (\ref{Xwave}), $U(\qq)= [ 2\pi e^2 /(\epsilon q) ] \exp(- q d)$ is the interlayer Coulomb energy,
where $\epsilon$ is the effective dielectric constant and $d$ is the vertical distance between W layer and Mo layer.

The effective mass $m_e$ and $m_h$ vary as a function of $\rr_0$.
{\it Ab initio} calculations show that the variation in $\mu$ (reduced mass) and $M$ (total mass) is only about $\pm 1 \%$,
which we will neglect for simplicity. 
For WS$_2$/MoS$_2$ bilayers in both AA and AB stackings,  
$m_e \approx 0.42 m_0$, $m_h \approx 0.34 m_0$ and $d \approx 6.15 $ \AA, where $m_0$ is the free electron mass.
The dielectric constant $\epsilon$ has a strong dependence on the environment.
Assuming that the heterobilayer is put on hexagonal boron nitride substrate and exposed to vacuum, 
$\epsilon$ is about $(5+1)/2=3$.
Equation (\ref{Xwave}) is solved numerically. We find that the binding energy $E_b$ is about 300 meV for the lowest energy exciton, of which $f(\kk)$ depends only on the magnitude of $\kk$. 
For comparison, the binding energy of the interlayer exciton in MoS$_2$/WSe$_2$
heterobilayer on sapphire substrate was determined to be 260meV in a previous experimental work.\cite{LiLJ_2014}
We approximate the exciton internal radius $a_X$ by the
root of mean square of the electron-hole separation.
$a_X$ can be estimated based on the wave function $f(\kk)$.
We obtain a numerical value of  $a_X$ about 1.3 nm. 
$a_X$ is much smaller compared to the moir\'e periodicity when the twist angle is smaller than 1$^{\circ}$.
Therefore, it is a good approximation to assume that the energy of the exciton follows the local band gap.

For a WSe$_2$/MoSe$_2$ bilayer, $m_e \approx 0.49 m_0$, $m_h \approx 0.35 m_0$ and $d \approx 6.47 $ \AA.
Still using $\epsilon\approx3$, we find $E_b$ is about 300 meV and $a_X$ is about 1.3 nm.

The lattice constant $a_0$ is about 3.19 \AA~ for WS$_2$ and 3.32 \AA~ for WSe$_2$.

\begin{widetext}

\section{Dependence of the optical spectrum on $\rr_0$}
\label{appB}
The global displacement $\rr_0$ just
leads to a spatial translation of the moir\'e potential. Therefore, the energy spectrum is independent of $\rr_0$. 
Here we further show that the optical spectrum is also independent of $\rr_0$.

The moir\'e potential has the form:
\begin{equation}
\Delta(\rr)=\sum_{j=1}^6 V_j \exp[i \bb_j \cdot (\rr+\rr_0')],
\end{equation}
where $\rr_0'=\rr_0 \times \hat{z} /\theta-\rr_0/2$.
The wave function at a finite $\rr_0$ is related to that at $\rr_0=0$ by applying a unitary transformation to the plane-wave state:
\begin{equation}
\begin{aligned}
|\chi\rangle=&e^{-i \qq_1\cdot\rr_0'}\Big( c_1 e^{i \qq_1\cdot\rr_0'}|\qq_1 \rangle + 
c_2 e^{i \qq_2\cdot\rr_0'}|\qq_2 \rangle +
c_3 e^{i \qq_3\cdot\rr_0'}|\qq_3 \rangle +...
\Big)\\
=& c_1 |\qq_1 \rangle + 
c_2 e^{i \bb_2\cdot\rr_0'}|\qq_2 \rangle +
c_3 e^{i \bb_3\cdot\rr_0'}|\qq_3 \rangle +...\\
\approx & c_1 |\qq_1 \rangle + 
c_2 e^{i \GG_T^{(2)}\cdot\rr_0}|\qq_2 \rangle +
c_3 e^{i \GG_T^{(3)}\cdot\rr_0}|\qq_3 \rangle +...,
\end{aligned}
\label{Utr}
\end{equation}
where $|\chi\rangle$ is the wave function at $\kappa$ point, and $...$ represents other components in $|\chi\rangle$.
In (\ref{Utr}), the approximation $\GG_T \approx \GG + \theta \hat{z} \times \GG  /2$ is used.
The phase factors $e^{i \GG_T^{(j)}\cdot\rr_0}$ capture the dependence on $\rr_0$, while the coefficient $c_j$ is independent of $\rr_0$.

The optical matrix element for $|\chi\rangle$ is:
\begin{equation}
\begin{aligned}
 &\frac{1}{\sqrt{\mathcal{A}}}\langle G |\hat{\jj}|\chi \rangle \\
=&\frac{1}{\sqrt{\mathcal{A}}}\sum_{n=1}^{3} \langle \qq_n |\chi\rangle \langle G |\hat{\jj}|\qq_n \rangle\\
=&c_1 \JJ_1 +\Big[c_2 e^{i \GG_T^{(2)}\cdot \rr_0}\Big]\Big[\JJ_2 e^{-i \GG_T^{(2)}\cdot \rr_0}\Big] +\Big[c_3 e^{i \GG_T^{(3)}\cdot \rr_0}\Big]\Big[\JJ_3 e^{-i \GG_T^{(3)}\cdot \rr_0}\Big]\\
=&c_1 \JJ_1 + c_2 \JJ_2 + c_3 \JJ_3,
\end{aligned}
\label{OME_rr0}
\end{equation}
where we have used Eq. (\ref{JJrr0}). 
Equation (\ref{OME_rr0}) proves that the optical spectrum is independent of $\rr_0$ for small but finite twist angle $\theta$.

\section{Results for $\text{WSe}_2$/$\text{MoSe}_2$ bilayer}
\label{appC}
The calculation for the WSe$_2$/MoSe$_2$ bilayer is parallel to that of the WS$_2$/MoS$_2$ bilayer, and the results are also quite similar.
For the WSe$_2$/MoSe$_2$ bilayer in AA stacking, $(V, \psi)=(11.8 \text{meV}, 79.5^{\circ})$ and $(|D_+|, |D_-|)=(0.22, 0.06)\mathcal{D}'$.
For the AB stacking, $(V, \psi)=(1.8 \text{meV}, 155.2^{\circ})$ and $(|D_+|, |D_-|)=(0.04, 0.12)\mathcal{D}'$.   Here $\mathcal{D}'$ measures the optical response of MoSe$_2$ intralayer excitation in AA stacking with zero twist angle and zero displacement.  $\mathcal{D}'$ is defined as $|\langle v \KK |\boldsymbol{\mathcal{J}}| c \KK \rangle |$, where both the conduction and valence states are located in the MoSe$_2$ layer. The numerical value of $\hbar \mathcal{D}'/e$ is about 4.43 eV$\cdot$\AA.

The potential energy, band structure, optical specturm and real space structure of interlayer excitons in WSe$_2$/MoSe$_2$ bilayer are summarized in Fig.~\ref{Fig:AA_spectrum_Se} and \ref{Fig:AB_spectrum_Se}.

\begin{figure}[h]
	\includegraphics[width=0.9\columnwidth]{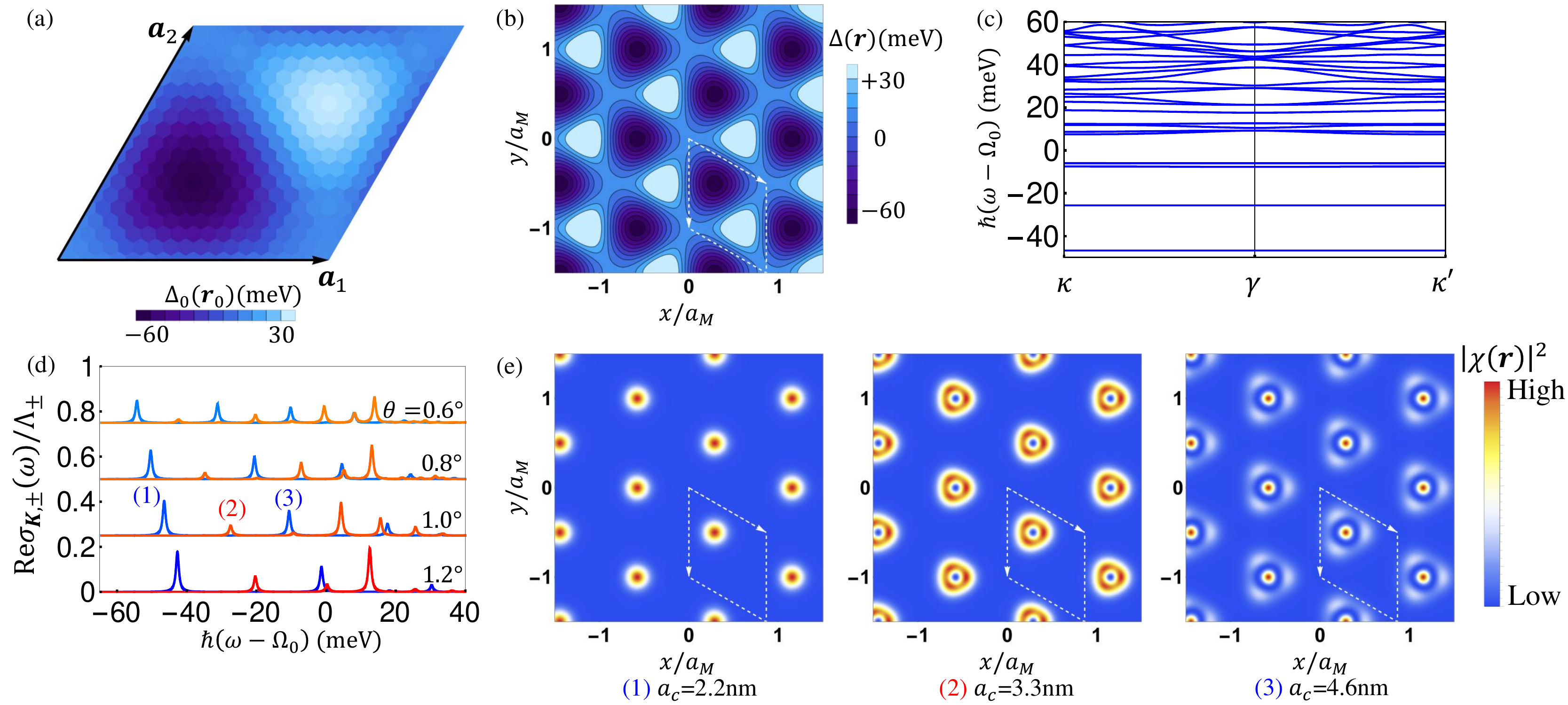}
	\caption{Results for WSe$_2$/MoSe$_2$ bilayer in AA stacking. }
	\label{Fig:AA_spectrum_Se}
\end{figure}

\begin{figure}[h]
	\includegraphics[width=0.9\columnwidth]{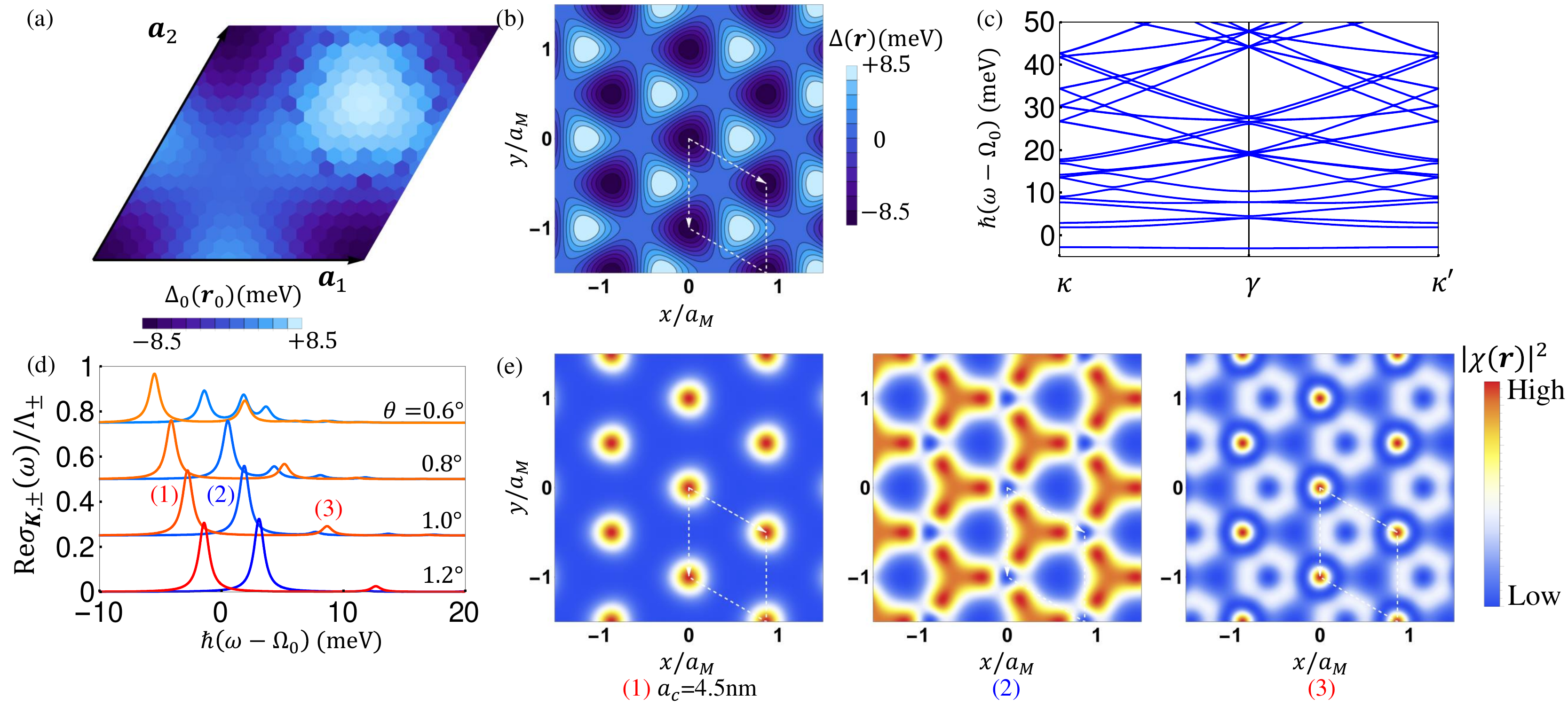}
	\caption{Results for WSe$_2$/MoSe$_2$ bilayer in AB stacking. }
	\label{Fig:AB_spectrum_Se}
\end{figure}

\end{widetext}

\newpage

\bibliographystyle{apsrev4-1}
\bibliography{refs}

\end{document}